\documentclass[12pt]{article}
\usepackage{amsbsy}

\newcommand{\sect}[1]{\setcounter{equation}{0}\section{#1}}
\newcommand{\eq}{\begin{equation}}
\newcommand{\eqa}{\begin{eqnarray}}
\newcommand{\en}{\end{equation}}
\newcommand{\ena}{\end{eqnarray}}
\newcommand{\enn}{\nonumber \end{equation}}

\def\sk{\vskip .4cm}
\def\noi{\noindent}
\def\om{\omega}

\def\ga{\gamma}
\def\Ga{\Gamma}

\let \part\partial

\def\unquarto{{1 \over 4}}

\def\unmezzo{{1 \over 2}}
\def\epsi{\varepsilon}
\def\we{\wedge}

\def\de{\delta}

\def\part{\partial}

\def\sk{\vskip .4cm}

\def\noi{\noindent}

\def\X0{X^0}

\def\om{\omega}

\def\ga{\gamma}

\def\unquarto{{1 \over 4}}
\def\unmezzo{{1 \over 2}}
\def\epsi{\varepsilon}
\def\epsibold{{\bf \epsilon}}

\def\we{\wedge}

\def\de{\delta}

\def\Dcal{{\cal D}}
\def\Rcal{{\cal R}}

\def\square{{\,\lower0.9pt\vbox{\hrule \hbox{\vrule height 0.2 cm
\hskip 0.2 cm \vrule height 0.2 cm}\hrule}\,}}

\def\epsilonbar{{\bar \epsilon}}

\def\Ftilde{\tilde F}

\def\psibar{\bar \psi}

\def\rhobar{\bar \rho}

\def\Om{\Omega}
\def\Qbar{\bar Q}
\def\neven{\bar n}
\def\reven{\bar r}
\def\seven{\bar s}
\def\nodd{\dot n}

\def\sodd{\dot s}

\def\Sigmabar{\overline \Sigma}

\def\Rbold{{\bf R}}
\def\Pbold{{\bf P}}
\def\Ombold{{\bf \Om}}
\def\onebold{{\bf 1}}
\def\epsibold{\boldsymbol {\epsilon}}

\def\Gammabold{{\bf \Gamma}}

\def\Gbold{{\bf G}}

\begin{document}

\begin{titlepage}
\rightline{ARC-17-02}
%\rightline{hep-th/9509031}
%\rightline{November 2011} 
\vskip 2em
\begin{center}
{\Large \bf  A locally supersymmetric $SO(10,2)$ invariant action for $D=12$ supergravity } \\[3em]

\vskip 0.5cm

{\bf
Leonardo Castellani}
\medskip

\vskip 0.5cm

{\sl Dipartimento di Scienze e Innovazione Tecnologica
\\Universit\`a del Piemonte Orientale, viale T. Michel 11, 15121 Alessandria, Italy\\ [.5em] INFN, Sezione di 
Torino, via P. Giuria 1, 10125 Torino, Italy\\ [.5em]
Arnold-Regge Center, via P. Giuria 1, 10125 Torino, Italy
}\\ [4em]
\end{center}

\begin{abstract}
\sk

We present an action for $N=1$ supergravity in $10+2$ dimensions, containing the gauge fields of the $OSp(1|64)$ superalgebra, i.e. one-forms $B^{(n)}$ with $n$=1,2,5,6,9,10 antisymmetric D=12 Lorentz indices and a Majorana gravitino $\psi$. The vielbein and spin connection correspond to $B^{(1)}$ and  $B^{(2)}$ respectively. The action is not gauge invariant under the full $OSp(1|64)$ superalgebra, but only under a subalgebra ${\tilde F}$ (containing the $F$ algebra $OSp(1|32)$), whose gauge fields are $B^{(2)}$, $B^{(6)}$, $B^{(10)}$ and the Weyl projected
Majorana gravitino ${1 \over 2} (1+\Gamma_{13}) \psi$. Supersymmetry transformations are therefore generated
by a Majorana-Weyl supercharge and, being part
of a gauge superalgebra, close off-shell. The action is simply $\int STr ({\bf R}^6 {\bf \Gamma})$
where ${\bf R}$ is the $OSp(1|64)$ curvature supermatrix two-form, and ${\bf \Gamma}$  is a constant supermatrix
involving $\Gamma_{13}$ and breaking $OSp(1|64)$ to its ${\tilde F}$ subalgebra. The action includes the usual Einstein-Hilbert term.

\end{abstract}

\vskip 5cm \noi \hrule \vskip .2cm \noi {\small
leonardo.castellani@uniupo.it}

\end{titlepage}

\newpage
\setcounter{page}{1}

\sect{Introduction}

Supergravity theories in dimensions greater than $D=11$ are believed to be inconsistent, since their reduction
to $D=4$ would produce more than $N=8$ supersymmetries, involving multiplets with spin $\ge 2$, and it is known that coupling of gravity with a finite number of higher spins is problematic. 

On the other hand a twelve dimensional theory with signature (10,2) avoids this difficulty, since fermions can be
both Majorana and Weyl in $D=10+2$, with 32 real components, and therefore giving rise to at most
eight supercharges when reduced to $D=4$. This fact has encouraged over the years various
attempts and proposals (\cite{Castellani1982} - \cite{Hewson1999}) for a twelve-dimensional field theory of supergravity.

A $D=10+2$ structure emerges also from string/brane theory, and has been named $F$-theory \cite{Vafa1996}.
The $OSp(1|32)$ superalgebra, a natural choice for the gauge algebra of a $D=10+2$ supergravity,
is also called $F$ algebra \cite{Hewson1999}.

Supergravity theories in various dimensions have been constructed using different approaches,
such as superspace (see for ex. \cite{superspace,vPF}), group geometric \cite{groupmanifold}, Chern-Simons \cite{Zanelli2005}, component (or ``Noether") techniques  (for ex. \cite{PvN1981,vPF}), or
combinations thereof  (for ex. \cite{Castellani2016}). 

Here we follow the ``gauge supergravity" road, inspired by Chern-Simons supergravities \cite{Zanelli2005}, and adapted to even dimensions via a higher curvature generalization of the
Mac Dowell-Mansouri (MDM) action \cite{MDM}. In the MDM framework (see also \cite{PvN2}) the fields of $N=1$, $D=4$ supergravity are
seen as  parts of the $OSp(1|4)$ superconnection, as originally proposed in \cite{Cham}, and enter the MDM action
only via the $OSp(1|4)$ curvature. The MDM action however is {\sl not} invariant under
$OSp(1|4)$ gauge transformations: indeed, besides the $OSp(1|4)$ curvature, it also contains
a constant matrix involving $\gamma_5$, that breaks $OSp(1|4)$ to its Lorentz subalgebra.
Since all supertranslations are broken, one may wonder how could this action be locally supersymmetric.
In fact one finds that it is supersymmetric in second order formalism (i.e. expressing the spin connection in terms of the vielbein and the gravitino  via the supertorsion constraint). But also remaining in first order formalism supersymmetry can be restored by
modifying the supersymmetry transformation law of the spin connection. The MDM action, after dispensing
with topological terms, becomes the action of (anti)de Sitter $N=1$ supergravity in $D=4$.
Note that supersymmetry here is not  {\sl gauge supersymmetry},  since it is not part of a superalgebra of transformations. There is no guarantee that the algebra of the ``restored supersymmetry"  closes off-shell, and indeed for this one needs auxiliary fields.

This being the state of affairs in $D=4$, can we do something similar in $D=12$ dimensions ? The present paper
provides an affirmative answer. The superalgebra $OSp(1|4)$ is replaced by
$OSp(1|64)$, the corresponding ``would be" gauge fields being one-forms 
$B^{(n)}$ with $n$=1,2,5,6,9,10 antisymmetric Lorentz indices and a Majorana gravitino $\psi$.
The vielbein and the spin connection are identified with $B^{(1)}$ and  $B^{(2)}$ respectively.
These one-forms are organized into an $OSp(1|64)$ connection, in an explicit 65 $\times$ 65
dimensional supermatrix representation. The $D=12$ action is constructed using exclusively
the $OSp(1|64)$ supermatrix curvature, and a constant matrix involving $\Gamma_{13}$.
This constant matrix ensures that the action is not topological (similarly to the MDM action) and breaks
$OSp(1|64)$ to a subalgebra $\Ftilde = OSp(1|32) \oplus Sp(32)$, under which the action is invariant\footnote{The $\Ftilde$ algebra contains the $F$ algebra: in fact the $F$ algebra is the $OSp(1|32)$ part 
of $\Ftilde$.}.
 Here part of the supersymmetry of
$OSp(1|64)$ survives, in contrast to the $D=4$ case. Supersymmetry is then a gauge symmetry, and 
closes off-shell. Twelve dimensional Lorentz symmetry $SO(10,2)$ is also part of the $\Ftilde$ gauge symmetry,
so that the action is $SO(10,2)$ invariant.

   Under the action of $\Ftilde$, the  $OSp(1|64)$ fields split into a gauge multiplet and a matter multiplet.
The gauge multiplet contains the $\Ftilde$ gauge fields: the spin connection $B^{(2)}$, a Majorana-Weyl gravitino $\psi_+$, and the other ``even" one-form fields
   $B^{(6)}$ and $B^{(10)}$. The matter multiplet contains the remaining $OSp(1|64)$ fields: the ``odd" one-form fields $B^{(1)}$ (the vielbein), $B^{(5)}$ and $B^{(9)}$, and a Majorana anti-Weyl gravitino $\psi_-$.
   
  Finally, let us emphasize that by ``$D=12$ supergravity" we really mean a field theory in $10+2$ dimensions,
  described by a geometrical action (the integral of a 12-form Lagrangian), containing the usual Einstein-Hilbert term
  and invariant under a gauge algebra that includes supersymmetry. This is much in the spirit of
  Chern-Simons supergravities in odd dimensions \cite{Troncoso1998}. The implications of the field equations
  for the correct counting of degrees of freedom are still to be investigated.
   
   The plan of the paper is as follows. Section 2  is a revisitation of the Mac Dowell-Mansouri action and its
   symmetries. Section 3 establishes notations for the $OSp(1|64)$ superalgebra, its connection and curvature
   components. In Section 4 the $D=12$, $N=1$ supergravity action is presented, and the proof of its invariance under 
   $\Ftilde$ gauge transformations is provided. Field transformation rules are given, and field equations are briefly discussed. Section 5 contains some conclusions and outlook. 
   
\sect{Mac Dowell-Mansouri action revisited}

This Section is just a review, and is included here because of its similarity to the $D=12$ construction. Most of it is taken from ref. \cite{Castellani2013}.

The Mac Dowell-Mansouri action \cite{MDM} is a $R^2$-type reformulation of (anti)de Sitter supergravity in $D=4$.
It is based on the supergroup $OSp(1|4)$, and the fields $V^a$ (vierbein), $\omega^{ab}$ (spin connection) 
and $\psi$ (gravitino\footnote{The gravitino $\psi$ is a Majorana spinor, i.e. $\psibar = \psi^T C$, where $C$ is the charge conjugation matrix.}) are 1-forms contained in the  $OSp(1|4)$ connection $\Ombold$, in a 5 $\times$ 5 supermatrix representation:

\eq
  \Ombold \equiv 
\left(
\begin{array}{cc}
  \Om &  \psi    \\
 \psibar  &  0   \\
\end{array}
\right), ~~~ \Om \equiv \unquarto \om^{ab} \ga_{ab} - {i \over 2} V^a \ga_a
    \label{Omdef4}  
  \en
The corresponding $OSp(1|4)$ curvature supermatrix is
 \eq
      \Rbold =  d \Ombold - \Ombold \we \Ombold~
  \equiv  \left(
\begin{array}{cc}
   R &  \Sigma    \\
 \Sigmabar  &  0   \\
\end{array}
\right) \label{Rdef4}
       \en
       \noindent and straightforward matrix algebra yields \footnote{we omit wedge products between forms.}:
   \eqa
    & & R = \unquarto R^{ab} \ga_{ab} - {i \over 2} R^a \ga_a  \label{defR4}\\
    & & \Sigma = d \psi - \unquarto \om^{ab} \ga_{ab} \psi + {i \over 2} V^a \ga_a \psi \label{defSigma4} \\
    & & \Sigmabar = d \psibar - \unquarto \psibar ~\om^{ab} \ga_{ab} + {i \over 2}  \psibar V^a \ga_a \\
        & & R^{ab} \equiv d \om^{ab} - \om^{a}_{~c} ~\om^{cb} + V^a V^b + {1 \over 2} \psibar \ga^{ab} \psi \\
     & & R^{a} \equiv d V^{a} - \om^{a}_{~b} V^{b}  -  {i \over 2} \psibar \ga^a \psi 
     \ena
 \noi We have also used the Fierz identity for $1$-form Majorana spinors:
  \eq
  \psi \psibar = \unquarto ( \psibar \ga^a \psi \ga_a - \unmezzo \psibar \ga^{ab} \psi \ga_{ab} )
  \en
\noi (to prove it, just multiply both sides by $\ga_c$ or $\ga_{cd}$ and take the  trace on spinor indices).  

The Mac Dowell-Mansouri action can be written in terms of the $OSp(1|4)$ curvature $\Rbold$ as:
 \eq
     S =  4 \int STr ( \Rbold \Gbold \Rbold \Gammabold)    \label{MDMaction}
    \en
    where $STr$ is the supertrace and $\Gbold, \Gammabold$ are the following constant matrices:
     \eq
       \Gammabold \equiv  \left(
\begin{array}{cc}
  i \ga_5 &  0    \\
 0 &  0   \\
\end{array}
\right), ~~~ \Gbold = \onebold + {\Gammabold^2 \over 2}= {1 \over 2} \left(
\begin{array}{cc}
 1 &  0    \\
 0 &  2   \\
\end{array}
\right) \label{Gammabold4}
       \en
\noi All boldface quantities are  5 $\times$ 5 supermatrices. 
Carrying out the supertrace, and then the spinor trace, leads to the familiar expression of the
MacDowell-Mansouri action:
\eq
 S = 2i \int Tr( R \we R \ga_5 + 2 \Sigma \we \Sigmabar \ga_5) = 2 \int \unquarto R^{ab} \we R^{cd} \epsi_{abcd} - 2i \Sigmabar \we \ga_5 \Sigma \label{MDMaction2}
  \en
  After inserting the curvature definitions the action takes the form
   \eq
    S = \int \Rcal^{ab} V^c V^d \epsi_{abcd} + 4 \rhobar \ga_a \ga_5 \psi V^a + \unmezzo (V^a V^b V^c V^d + 2 \psibar \ga^{ab} \psi V^c V^d ) \epsilon_{abcd}  \label{adsSG}
    \en
  with
   \eq
    \Rcal^{ab} \equiv d \om^{ab} - \om^{a}_{~c} ~\om^{cb} , ~~\rho \equiv d \psi - \unquarto \om^{ab} \ga_{ab} \psi \equiv \Dcal \psi
     \label{RLordef}
     \en
   We have dropped the topological term $\Rcal^{ab} \Rcal^{cd} \epsilon_{abcd}$ (Euler form), and used 
   the gravitino Bianchi identity 
    \eq
     \Dcal \rho = - \unquarto \Rcal^{ab} \ga_{ab} \psi
   \en
   and the gamma matrix identity $2 \ga_{ab} \ga_5 = i \epsilon_{abcd} \ga^{cd}$  to recognize that $\unmezzo \Rcal^{ab} \psibar \ga^{cd} \psi \epsilon_{abcd} - 4 i \rhobar \ga_5 \rho$
    is a total derivative. The action (\ref{adsSG}) describes $N=1$, $D=4$ anti-De Sitter supergravity,
    the last term being the supersymmetric cosmological term. After rescaling the vielbein and the gravitino as
    $V^a \rightarrow \lambda V^a$, $\psi  \rightarrow \sqrt{\lambda} \psi$ and dividing the action by $\lambda^2$, the usual (Minkowski)
    $N=1$, $D=4$ supergravity is retrieved by taking the limit $\lambda \rightarrow 0$. This corresponds to the
Inon\"u-Wigner contraction of $OSp(1|4)$ to the superPoincar\'e group.
\sk
\noi {\bf Invariances}
\sk
 \noi As is well known, the action (\ref{MDMaction}), although a bilinear in the $OSp(1|4)$ curvature, is {\it not} invariant under the 
  $OSp(1|4)$ gauge transformations:
   \eq
   \de_{\epsibold} \Ombold = d \epsibold - \Ombold \epsibold + \epsibold \Ombold~ \Longrightarrow~  \de_{\epsibold} \Rbold =  - \Rbold \epsibold + \epsibold \Rbold   \label{Rgauge4}
       \en
       where $\epsibold$ is the $OSp(1|4)$ gauge parameter:
        \eq
         \epsibold \equiv 
         \left(
\begin{array}{cc}
 \unquarto  \epsi^{ab} \ga_{ab} - {i \over 2} \epsi^a \ga_a &  \epsilon   \\
 \epsilonbar  &  0   \\
\end{array}
\right)   \label{gaugeparam4}
 \en
 In fact it is not a Yang-Mills action (involving the exterior product of $\Rbold$ with its Hodge dual), nor a topological action  $\int STr(\Rbold \Rbold)$: the constant supermatrices $\Gbold$ and  $\Gammabold$ ruin the $OSp(1|4)$ gauge invariance, and break it to its Lorentz subgroup.  Indeed the gauge variation of the action (\ref{MDMaction})
 \eq
\delta S = 4 ~ \int STr (\Rbold [\Gbold,\epsibold] \Rbold \Gammabold + \Rbold \Gbold \Rbold [\Gammabold,\epsibold] ) \label{gaugevariation4}
 \en
  vanishes when $\epsibold$ commutes with $\Gammabold$ (and therefore with $\Gbold$), and this happens
  only when $\epsibold$ in (\ref{gaugeparam4}) has $\epsi^{a} = \epsilon =0$, so that only Lorentz rotations leave the action invariant. 
  
  Specializing the gauge parameter $\epsibold$ to describe supersymmetry variations (i.e. only $\epsilon \not= 0$ in (\ref{gaugeparam4})),
  eq. (\ref{gaugevariation4}) yields the supersymmetry variation of the Mac Dowell-Mansouri action: 
  \eqa
  & & \delta_{susy}S= 2i ~\int (\epsilonbar [\gamma_5 , R] \Sigma + \Sigmabar [ \gamma_5 , R ] \epsilon) \label{susyvariation} \\
  & & ~~~~~~~~ =  -4 \int R^a \Sigmabar \ga_a \ga_5 \epsilon
  \ena
 with $R$ defined in (\ref{defR4}). This variation is proportional to the torsion $R^a$, since only
 $R^a \gamma_a$ in $R$ has a nonzero commutator with $\gamma_5$.
 \
  Therefore in second-order 
 formalism, i.e. using the torsion constraint $R^a=0$ to express $\omega^{ab}$ in terms of $V^a$ and $\psi$,  the action is indeed supersymmetric. Another way to recover supersymmetry is by modifying the
 supersymmetry variation of the spin connection, see for ex. \cite{PvN2}. In both cases supersymmetry is
 not part of a gauge superalgebra: off-shell closure of the supersymmetry transformations is not
 automatic, and indeed necessitates the introduction of auxiliary fields.
 \sk
 We shall now jump to 12 dimensions, and write a geometrical $\Rbold^6$-type action that resembles
 the $\Rbold^2$-type Mac Dowell-Mansouri action of $D=4$. The action will be invariant under the $\Ftilde$ subalgebra
 of $OSp(1|64)$. Contrary to the $D=4$ case, $N=1$ supersymmetry (with a Majorana-Weyl 
 supercharge) survives as part of
 the $\Ftilde$ algebra, and closes off-shell. 
 
 \sect{$OSp(1|64)$ connection and curvature}

{\bf The algebra }
\sk

\noi The $OSp(1|64)$ superalgebra has the following structure:

\eqa
& & [Z_{(n)},Z_{(m)}]= c_{(n)(m)}^{(r)} Z_{(r)} \label{ZZcomm}\\
 & & \{ \Qbar_{\alpha}, \Qbar_{\beta} \} =  {1 \over 16}~ {(-1)^n \over n!}(C\Gamma^{(n)})_{\alpha\beta} Z_{(n)} \label{QQcomm} \\
  & & [Z_{(n)},\Qbar_{\alpha}]={1 \over 2} ~(\Gamma_{(n)})^{\beta}_{~\alpha} \Qbar_\beta \label{ZQcomm}
\ena
where the bosonic generators $Z_{(n)}$ with $n$=1,2,5,6,9,10 have $n$ antisymmetrized Lorentz indices and are dual to the one-forms $B^{(n)}$, and
the Majorana supersymmetry charge $\Qbar$ is dual to the Majorana gravitino one-form $\psi$. 
In (\ref{QQcomm}) and (\ref{ZQcomm}) the D=12 gamma matrices  $\Gamma^{(n)}$ with $n$ antisymmetrized indices appear as structure constants, and $C$ is the charge conjugation matrix. The structure constants
$c_{(n)(m)}^{(r)}$ in (\ref{ZZcomm}) are obtained from the commutation relations of the $\Gamma^{(n)}$,
and are given in Appendix A.
Repeated $(n)$ indices are summed on $n$=1,2,5,6,9,10.
\sk\sk
\noi {\bf The 65 $\times$ 65 supermatrix representation}
\sk

\noi Gamma matrices in $D=12$ have a  $64 \times 64$ matrix representation 
and the superalgebra above can be realized by a
65 $\times$ 65 supermatrix representation:

  \eq
  Z_{(n)} = \left(
\begin{array}{cc}
 {1 \over 2} \Gamma_{(n)} &  0    \\
 0  &  0   \\
\end{array}
\right)  , ~~\Qbar_\alpha= Q^\beta C_{\beta\alpha} =  \left(
\begin{array}{cc}
0 &  \delta^\rho_\alpha    \\
 C_{\sigma\alpha} &  0   \\
\end{array} \right)\label{representation}  
  \en
  To verify the anticommutations
  (\ref{QQcomm}), one needs the identity 
   \eq
   \delta^\rho_\alpha C_{\sigma\beta} +   \delta^\rho_\beta C_{\sigma\alpha} = {1 \over 32} {(-1)^n \over n!}    (C\Gamma_{(n)})_{\alpha\beta} (\Gamma^{(n)})^\rho_{~\sigma} 
       \en
 deducible from the Fierz identity (\ref{2psiFierz}) by factoring out the two spinor Majorana one-forms.
\sk\sk
\noi {\bf Connection and curvature }
\sk

\noi The $1$-form $OSp(1|64)$-connection  is given by
\eq
\Ombold = {1 \over n!} B^{(n)} Z_{(n)} + \Qbar_\alpha \psi^\alpha
\en
In the 65 $\times$ 65 supermatrix representation:
 \eq
  \Ombold \equiv 
\left(
\begin{array}{cc}
  \Om &  \psi    \\
- \psibar  &  0   \\
\end{array}
\right), ~~~ \Om = {1 \over 2 \cdot n!}  B^{(n)} \Gamma_{(n)}  
    \label{Omdef}  
  \en
 The corresponding curvature supermatrix 2-form is
 \eq
      \Rbold =  d \Ombold - \Ombold \we \Ombold~
  \equiv  \left(
\begin{array}{cc}
   R &  \Sigma    \\
- \Sigmabar  &  0   \\
\end{array}
\right) \label{Rdef}
       \en
 with\footnote{we omit wedge products between forms}
   \eqa
    & & R = d \Omega - \Omega \Omega + \psi \psibar \equiv  {1 \over 2 \cdot n!} R^{(n)} \Gamma_{(n)}  \label{defR}\\
    & & \Sigma = d \psi - \Omega \psi = d \psi -  {1 \over 2 \cdot n!}  B^{(n)} \Gamma_{(n)} \psi \label{defSigma} \\
    & & \Sigmabar = d \psibar - \psibar \Omega= d \psibar - {1 \over 2 \cdot n!}  \psibar B^{(n)} \Gamma_{(n)}
        \ena
    \noi The bosonic curvature components $R^{(n)}$ are easily determined:
     \eqa
     & & R^{(n)} = d B^{(n)} + {1 \over 2} c_{(m)(r)}^{(n)} B^{(m)} B^{(r)} - {(-1)^n \over 32} \psibar \Gamma^{(n)} \psi
     \ena
 \noi where we used the Fierz identity for $1$-form Majorana  spinors in (\ref{2psiFierz}). 
For example for $R^{(2)}$ we find:
 \eqa
 & & R^{ab}=dB^{ab} - B^{ac} B_c^{~b} - V^a V^b -  {1 \over 32} \psibar \Gamma^{ab} \psi ~~~~~~~~~~~~~~\nonumber \\
& & ~~~~~ - {1 \over 4!} B^{c_1 - c_4 a} B_{c_1 - c_4}^{~~~~~b} + {1 \over 5!} B^{c_1 - c_5 a} B_{c_1 - c_5}^{~~~~~b} - {1 \over 8!} B^{c_1 - c_8 a} B_{c_1 - c_8}^{~~~~~b} + {1 \over 9!} B^{c_1 - c_9 a} B_{c_1 - c_9}^{~~~~~b}
\nonumber \\
 \ena
 
 \sect{The $D=12$, $N=1$ supergravity action}
 
    An $\Ftilde$-invariant action for the fields contained in the $OSp(1|64)$ connection  can be written as the integral of a Lagrangian 12-form, using  the $OSp(1|64)$ curvature supermatrix $\Rbold$:
    \eq
     S =   \int STr ( \Rbold^6 \Gammabold)    \label{action}
    \en
    where $STr$ is the supertrace and $\Gammabold$ is the constant matrix
     \eq
 \Gammabold \equiv  \left(
\begin{array}{cc}
  \Gamma_{13 }&  0    \\
 0 &  1  \\
\end{array}
\right)\label{Gammabold}
       \en
  \sk
 \noi  {\bf Invariances}
  \sk

 \noi Under the $OSp(1|64)$
  gauge transformations:
   \eq
   \de_{\epsibold} \Ombold = d \epsibold - \Ombold \epsibold + \epsibold \Ombold ~ \Longrightarrow~  \de_{\epsibold} \Rbold =  - \Rbold \epsibold + \epsibold \Rbold     \label{Omgaugevariation}
       \en
       where $\epsibold$ is  the $OSp(1|64)$ gauge parameter:
       \eq
         \epsibold \equiv 
         \left(
\begin{array}{ll}
{1 \over 2 \cdot n!}   \epsi^{(n)} \Gamma_{(n)}  &  \epsilon   \\
- \epsilonbar  &  0   \\
\end{array}
\right)   \label{gaugeparam}
 \en
  the action (\ref{action}) varies as
 \eq
\delta S =  \int STr (\Rbold^6  [\Gammabold,\epsibold] ) \label{Sgaugevariation}
 \en
 Now
 \eq
  [\Gammabold,\epsibold] =   \left(
\begin{array}{cc}
{1 \over 2 \cdot n!}   \epsi^{(n)}  [\Gamma_{13} ,  \Gamma_{(n)} ] &  (\Gamma_{13} - 1) \epsilon   \\
 \epsilonbar (\Gamma_{13} - 1)  &  0   \\
\end{array}
\right)  \label{gammaepsiloncomm}
\en
 \noi vanishes when the gauge parameters $\epsi^{(n)}$ have even $n$, so that $\Gamma_{(n)}$ commutes
 with $\Gamma_{13}$, and $\epsilon$ is
 a Weyl spinor (besides being Majorana), i.e. $\Gamma_{13} \epsilon = \epsilon$.
 These restrictions on the gauge parameters determine a subalgebra of $OSp(1|64)$, generated by
 $Z_{(2)}$,  $Z_{(6)}$, $Z_{(10)}$,  and  $\Qbar_{\alpha} P_{+}$, where $P_{+} = {1 \over 2} (1+\Gamma_{13})$. These generators close on the $\Ftilde$ = $OSp(1|32) \oplus Sp(32)$ subalgebra
 of $OSp(1|64)$. 
 \sk
 \noi {\bf Note:} the analogous action in four dimensions  $\int {\rm STr} ( \Rbold^2 \Gammabold) $, where 
 $\Gammabold$ is as in (\ref{Gammabold}) with $\Gamma_{13}$ substituted by $\gamma_5$, is not invariant under (chiral) gauge supersymmetry since the
 supersymmetry parameter cannot be Majorana-Weyl in $D=3+1$.
\sk
 
\noi The $F$ algebra is contained in $\Ftilde$ = $OSp(1|32) \oplus Sp(32)$, and corresponds to the $OSp(1|32)$ term in the direct sum. Its generators are
given by $Z_{(2)}^+ \equiv Z_{(2)} \Pbold_+$ (66), $Z_{(6)}^+ \equiv Z_{(6)} \Pbold_+$ (462, selfdual) and $\Qbar_\alpha^+ \equiv \Qbar_\alpha \Pbold_+$ (32), where the number of the generators is given between parentheses, and
\eq
\Pbold_{\pm} \equiv  \left(
\begin{array}{cc}
  {1 \over 2} (1 \pm\Gamma_{13 }) &  0    \\
 0 &  0  \\
\end{array}
\right)
\en
The $Sp(32)$ part
of $\Ftilde$ is generated by $Z_{(2)}^- \equiv Z_{(2)} \Pbold_-$ (66) and $Z_{(6)}^- \equiv Z_{(6)} \Pbold_-$ (462, antiselfdual). The $Z_{(10)}$ generators 
of $\Ftilde$ are absorbed into $Z_{(2)} \Pbold_\pm$ since $\Gamma_{(2)} \Gamma_{13} $ is proportional to
$\Gamma_{(10)}$. The Weyl supersymmetry charges $\Qbar^+_\alpha$ close on $Z_{(2)}^+$ and $Z_{(6)}^+$.
The $\Ftilde$ commutations in terms of the $OSp(1|32)$ and  $Sp(32)$ generators are given
in Appendix B.
\sk\sk
 \noi{\bf $\Ftilde$ transformation laws}
 \sk
\noi  Restricting the gauge parameter $\epsibold$ to the $\Ftilde$ subalgebra as described above, from (\ref{Omgaugevariation}) we deduce the $\Ftilde$  transformation laws on the fields $B^{(n)}$ ($n$=1,2,5,6,9,10) and $\psi$:
 \eqa
 & &  \delta B^{({\bar n})} = d \epsi^{(\neven)} + { \neven ! \over \reven ! \seven !}  c^{(\neven)}_{(\reven)(\seven)} \epsi^{(\reven)} B^{(\seven)} - {1 \over 16} \psibar \Gamma^{(\neven)}    \epsilon  \equiv D^{\Ftilde}   \epsi^{(\neven)} - {1 \over 16}  {\bar \psi_+} \Gamma^{(\neven)} \epsilon \\
 & &  \delta B^{(\nodd)} = { \nodd ! \over \reven ! \sodd !}  c^{(\nodd)}_{(\reven)(\sodd)} \epsi^{(\reven)} B^{(\sodd)} + {1 \over 16} {\bar \psi_-}  \Gamma^{(\nodd)}    \epsilon     \\
 & & \delta \psi_{+} = d \epsilon - {1 \over 2 \cdot \neven !} B^{(\neven)} \Gamma_{(\neven)} \epsilon + 
  {1 \over 2 \cdot \neven !} \epsi^{(\neven)} \Gamma_{(\neven)} \psi_{+} \equiv D^{\Ftilde} \epsilon +
   {1 \over 2 \cdot \neven !} \epsi^{(\neven)} \Gamma_{(\neven)} \psi_{+} ~~~~~\\
  & & \delta \psi_{-} = - {1 \over 2 \cdot \nodd !} B^{(\nodd)} \Gamma_{(\nodd)} \epsilon + 
  {1 \over 2 \cdot \neven !} \epsi^{(\neven)} \Gamma_{(\neven)} \psi_{-} 
 \ena
 where $\epsilon$ is the Weyl projected supersymmetry parameter, indices with a bar run on even values, and indices with a dot on odd values, i.e.
 $\neven = 2,6,10$ and $\nodd = 1,5,9$. Moreover $\psi_+$ and $\psi_-$ are respectively Weyl and anti-Weyl projections of the Majorana gravitino, i.e. $\psi_\pm = P_\pm \psi$. 
 
 Thus we see that the $\Ftilde$ gauge fields $B^{\neven}$, $\psi_+$ transform with the
 $\Ftilde$ covariant derivative of the gauge parameters, whereas the ``matter fields"
 $B^{\nodd}$, $\psi_-$ transform homogeneously. These last include the vielbein $V = B^{(1)}$. Note also
 that gauge and matter fields do not mix, separating into a gauge and a matter multiplet under 
 $\Ftilde$ transformations.
 \sk
\sk
\noi {\bf Einstein-Hilbert term}
\sk
\noi The Lagrangian 12-form $STr(\Rbold^6 \Gammabold)$ contains the usual Einstein-Hilbert term
 \eq
  \Rcal^{a_1a_2} \wedge V^{a_3} \wedge \cdots \wedge V^{a_{12}} \epsi_{a_1 - a_{12}} \label{EHterm}
  \en
  where $\Rcal$ is the Lorentz curvature
  \eq
  \Rcal^{a_1a_2} = dB^{a_1a_2} -B^{a_1}_{~~c}  \wedge B^{ca_2}
   \en
   Indeed $\Rbold^6 \Gammabold$ includes a term (in the upper left corner of the supermatrix):
   \eq
   \Rcal^{a_1a_2} \wedge V^{a_3} \wedge V^{a_4} \we \cdots \we V^{a_{11} }\we V^{a_{12}} \Gamma_{a_1a_2} \Gamma_{a_3a_4} \cdots \Gamma_{a_{11}a_{12}} \Gamma_{13}
   \en
   Tracing on spinor indices produces the EH term (\ref{EHterm}), since $Tr[\Gamma_{a_1a_2} \Gamma_{a_3a_4} \cdots \Gamma_{a_{11}a_{12}} \Gamma_{13}]$ is proportional to $\epsi_{a_1 - a_{12}}$.

  \sk \sk
 \noi{\bf Field equations}
 \sk
\noi They are obtained by varying the $OSp(1|64)$ superconnection $\Ombold$ in (\ref{action}), yielding:
\eq
\delta S = \int STr[  \{ (D \delta \Ombold) \Rbold^5 \} \Gammabold]=0
\en
where we have set for short
\eq
 \{ (D \delta \Ombold) \Rbold^5 \} \equiv (D  \delta \Ombold) \Rbold^5 + \Rbold (D  \delta \Ombold) \Rbold^4 + \cdots + \Rbold^5 (D  \delta \Ombold)
 \en
 and $D$ is the $OSp(1|64)$ covariant derivative. Integrating by parts and using
 \eq
 D \Gammabold = - \Ombold \Gammabold + \Gammabold \Ombold= 
  \left(
\begin{array}{cc}
  [\Gamma_{13} ,  \Omega] &  (\Gamma_{13} - 1) \psi   \\
 \psibar (\Gamma_{13} - 1)  &  0   \\
\end{array}
\right) 
\en
leads to
\eq
\delta S = -2 \int STr [  \{  \delta \Ombold \Rbold^5 \}  
\left(
\begin{array}{cc}
 \Omega^{(odd)} \Gamma_{13} &   \psi_-  \\
 {\bar \psi_-}  &  0   \\
\end{array}
\right) ] =0 \label{fieldeq}
\en
where $\Omega^{(odd)}$ contains only odd gamma matrices. Further analysis of the field equations
(\ref{fieldeq}), of their solutions and of possible ``compactifications" relating to lower
dimensional supergravities, notably $D=10$, 2b supergravity, will be reported in a subsequent paper.
 
\sect{Conclusions}

The $D=12$ supergravity action we have presented here 
is made out of the fields contained in the $OSp(1|64)$ connection, but is invariant
only under a subalgebra $\Ftilde$ of  $OSp(1|64)$. This closely resembles what happens for the
Mac Dowell-Mansouri action in $D=4$, where the supergavity fields are organized in 
a $OSp(1|4)$ connection, but the action itself is invariant only under the Lorentz subalgebra.
In the $D=12$ case, however, supersymmetry is part of the invariance subalgebra of the action, since a Majorana-Weyl
supercharge is included in $\Ftilde$. This supercharge can give rise to at most $N=8$ supersymmetries once the action is reduced to four dimensions.  

We could try to export the constructive procedure adopted in this paper to the case of $D=9+1$ supergravity. 
The relevant ``starting" superalgebra is $OSp(1|32)$,
generated by $Z^{(n)}$ with $n$=(1,2,5,6,9,10) antisymmetric $D=10$ Lorentz indices, and a 
Majorana charge. The same argument of Section 4 would lead to an action $\int STr(\Rbold^5 \Gammabold)$ 
where $\Gammabold$ involves $\Gamma_{11}$.  This action however is invariant only under the bosonic subalgebra of  
$OSp(1|32)$ generated by the $Z^{(n)}$ with $n$=(2,6,10), and not under the supersymmetry generated by a Weyl-projected Majorana charge.
Indeed, contrary to the $D=10+2$ case, $\Ga_{11} \epsilon = \epsilon$ implies
$\epsilonbar \Ga_{11} = - \epsilonbar$ (note the minus sign), so that
the analogue of the commutator (\ref{gammaepsiloncomm}) does not vanish for 
supersymmetry variations.

\section*{Acknowledgements}

It is a pleasure to acknowledge useful discussions with my colleagues and friends Laura Andrianopoli,  Paolo Aschieri, Riccardo D' Auria, Pietro Fr\'e, Pietro Antonio Grassi, Mario Trigiante, Toine van Proeyen, and Jorge Zanelli.

\appendix

\sect{D=12 $\Gamma$ matrices, $Sp(64)$ algebra and a Fierz identity}

\noi {\bf Clifford algebra}
\eq
\{ \Gamma_a,\Gamma_b \}= 2 \eta_{ab}, ~~~~\eta_{ab} = (+1,+1, -1, \cdots, -1),~~~~\Gamma_{13}=\Gamma_1 \cdots \Gamma_{12}, ~~~~\Gamma_{13}^2 = 1
\en
\noi The antisymmetrizations of $n$ Lorentz indices in $\Gamma_{(n)}$ have weight 1. For ex.
$\Gamma_{ab} \equiv {1 \over 2} [\Gamma_a,\Gamma_b]$. The $D=12$ charge conjugation matrix $C$ satisfies
\eq
\Gamma_a^T= - C \Gamma_a C^{-1} ,~~~C^T=-C
\en
so that $C\Gamma_{(n)}$ with $n$=1,2,5,6,9,10 are symmetric, and the remaining $C\Gamma_{(n)}$
with $n$= 3,4,7,8,11,12 are antisymmetric. 
\sk
\noi {\bf Duality relations}:
\eq
\Gamma_{a_1 \cdots a_n} = (-1)^{n(n-1)\over 2} {1 \over (12-n)!} \epsi_{a_1 \cdots a_n}^{~~~~~~a_{n+1} \cdots a_{12}} ~\Gamma_{a_{n+1} \cdots a_{12}} \Gamma_{13}
\en
\noi or, in a short-hand notation:
\eq
\Gamma_{(n)} = (-1)^{n(n-1)\over 2} {1 \over (12-n)!} \epsi_{(n)}^{~~(12-n)} ~\Gamma_{(12-n)} \Gamma_{13}
\en
Defining the Weyl projectors
\eq
P_\pm = {1 \over 2} (1 \pm \Gamma_{13})
\en
one finds for example
\eqa
& & \Gamma_{(10)}P_\pm = \mp {1 \over 2} \epsi_{(10)}^{~~~(2)} ~\Gamma_{(2)} P_\pm \label{Gamma10P}\\
& & \Gamma_{(6)}P_\pm = \mp {1 \over 6!} \epsi_{(6)}^{~~~(6)} ~\Gamma_{(6)} P_\pm \label{Gamma6P}
\ena
\noi {\bf The algebra of $\Gamma_{(n)}$}
\eq
[\Gamma_{(n)},\Gamma_{(m)}]= 2 c_{(n)(m)}^{(r)} ~\Gamma_{(r)} \label{Sp64}
\en
The $\Gamma_{(n)}$ with $n$=1,2,5,6,9,10 generate the $Sp(64)$ algebra. This algebra contains
$Sp(32) \oplus Sp(32)$, generated by the even $\Gamma$'s : $\Gamma_{(2)}$, $\Gamma_{(6)}$ and
$\Gamma_{(10)}$. The two $Sp(32)$ are separately generated by $\Gamma_{(2)}P_+$, $\Gamma_{(6)}P_+$,
and by $\Gamma_{(2)}P_-$, $\Gamma_{(6)}P_-$. Note that $\Gamma_{(10)} P_\pm$ is proportional to
$\Gamma_{(2)} P_\pm$ because of (\ref{Gamma10P}) and that $\Gamma_{(6)} P_+$ and 
$\Gamma_{(6)} P_-$   are respectively selfdual and anti-selfdual. 
\sk
\noi The explicit form of the commutations (\ref{Sp64}) is:
\eq
[\Gamma^{a_1- a_n},\Ga_{b_1 - b_m}] = 2 \sum_{k=0}^{min[m,n]} C_{n,m}^{n+m-2k} \delta^{[a_1 - a_k}_{[b_1 - b_k}
\Gamma^{a_{k+1} - a_n]}_{~~~~~~~~~b_{k+1} - b_m]}
\en
where the nonvanishing structure constants $C_{n,m}^{r}$ are given in the following Table:
\sk

\begin{tabular}{|c|c|c|c||c|c|c|c||c|c|c|c|} \hline $n$ & $m$ &$ r $& $C_{n,m}^{r}$& $n$ & $m$ &$ r $& $C_{n,m}^{r}$ & $n$ & $m$ &$ r $& $C_{n,m}^{r}$  \\\hline 1 & 1 & 2 & 1 & 5 & 5 & 2 & 5 $\cdot$ 5!  & 6 & 10 & 10 & 20 $\cdot$ 6! \\\hline 1 & 2 & 1 & 2 & 5 & 6 & 9 & 30 & 6 & 10 & 6 & -9!/2 \\\hline 1 & 5 & 6 & 1 & 5 & 6 & 5 & - 10 $\cdot$ 5! & 9 & 9 & 10 & 21 $\cdot$9!/20 \\\hline 1 & 6 & 5 & 6 & 5 & 6 & 1 & 6! & 9 & 9 & 6 & -2 $\cdot$ 7 $\cdot$ 9! \\\hline 1 & 9 & 10 & 1 & 5 & 9 & 10 & -6! & 9 & 9 & 2 & 9 $\cdot$ 9! \\\hline 1 & 10 & 9 & 10 & 5 & 9 & 6 & 3 $\cdot$ 7! & 9 & 10 & 9 & 21 $\cdot$ 9!/2 \\\hline 2 & 2 & 2 & -4 & 5 & 10 & 9 & -10 $\cdot$ 6! & 9 & 10 & 5 & - 6 $\cdot$ 10! \\\hline 2 & 5 & 5 & -10 & 5 & 10 & 5 & 6 $\cdot$ 7! & 9 & 10 & 1 & 10! \\\hline 2 & 6 & 6 & -12 & 6 & 6 & 10 & -36 & 10 & 10 & 10 & -21 $\cdot$ 9! \\\hline 2 & 9 & 9 & -18 & 6 & 6 & 6 & 20 $\cdot$ 5! & 10 & 10 & 6 & 20 $\cdot$ 10! \\\hline 2 & 10 & 10 & -20 & 6 & 6 & 2 & -6 $\cdot$ 6! & 10 & 10 & 2 & -10 $\cdot$ 10!  \\\hline 5 & 5 & 10 & 1 & 6 & 9 & 9 & 8!/4 &  &  &  &  \\\hline 5 & 5 & 6 & -200 & 6 & 9 & 5 & -9!/4 &  &  &  & \\\hline \end{tabular}

\sk
\noi {\bf Fierz identity }
\sk
\noi Majorana spinor one-forms ($\psibar = \psi C$) satisfy the Fierz identity:
 \eq
 \psi \psibar = - {1 \over 64} \sum_n {(-1)^n \over n!} \psibar \Gamma^{(n)} \psi \Gamma_{(n)}~~~n=1,2,5,6,9,10
 \label{2psiFierz}
  \en
 where for example $\psibar \Gamma^{(2)} \psi \Gamma_{(2)}$ is defined by $\psibar \Gamma^{a_1a_2} \psi \Gamma_{a_1a_2}$ etc. To prove it, multiply by $\Gamma_{(m)}$
 and take the trace of both sides of (\ref{2psiFierz}).
 
 \sect{ The $\Ftilde$ superalgebra}
\noi The $\Ftilde$ generators satisfy the following (anti)commutation relations:
 \eqa
& & [Z_{(2)}^\pm,Z_{(2)}^\pm]= c_{(2)(2)}^{(2)} Z_{(2)}^\pm \\
& & [Z_{(2)}^\pm,Z_{(6)}^\pm]= c_{(2)(2)}^{(6)} Z_{(6)}^\pm \\
& & [Z_{(6)}^\pm,Z_{(6)}^\pm]= [c_{(6)(6)}^{(2)} - {1 \over 2}  c_{(6)(6)}^{(10)} \epsi_{(10)}^{~~(2)} ] Z_{(2)}^\pm
+ c_{(6)(6)}^{(6)} Z_{(6)}^\pm\\
& &[Z_{(r)}^\pm,Z_{(s)}^\mp]=0,~~~~r,s=2,6\\
 & & \{ \Qbar_{\alpha}^+, \Qbar_{\beta}^+ \} =  {1 \over 16 \cdot n!}(C\Gamma^{(n)}P_+)_{\alpha\beta} Z_{(n)}^+ ,~~~~~n=2,6 \\
  & & [Z_{(n)}^+,\Qbar_{\alpha}^+]={1 \over 2} ~(\Gamma_{(n)} P_+)^{\beta}_{~\alpha} \Qbar_\beta^+,~~~~n=2,6 \\
    & & [Z_{(n)}^-,\Qbar_{\alpha}^+]=0,~~~~n=2,6
\ena
The generators with a + superscript belong to $OSp(1|32)$, those with a - superscript to $Sp(32)$.


\begin{thebibliography}{99}
 
 \bibitem{Castellani1982} 
  L.~Castellani, P.~Fr\'e, F.~Giani, K.~Pilch and P.~van Nieuwenhuizen,
  ``Beyond $d=11$ Supergravity and Cartan Integrable Systems,''
  Phys.\ Rev.\ D {\bf 26}, 1481 (1982).
    %%CITATION = doi:10.1103/PhysRevD.26.1481;%%
    
    \bibitem{Kutasov1996} 
  D.~Kutasov and E.~J.~Martinec,
  ``New principles for string / membrane unification,''
  Nucl.\ Phys.\ B {\bf 477}, 652 (1996)
  [hep-th/9602049].
  %%CITATION = doi:10.1016/0550-3213(96)00302-1;%%
  
    \bibitem{Bars1996bis} 
  I.~Bars,
  ``Supersymmetry, p-brane duality and hidden space-time dimensions,''
  Phys.\ Rev.\ D {\bf 54}, 5203 (1996)
    [hep-th/9604139].
  %%CITATION = doi:10.1103/PhysRevD.54.5203;%%
ì
   \bibitem{Bars1996} 
  I.~Bars,
  ``S theory,''
  Phys.\ Rev.\ D {\bf 55}, 2373 (1997)
    [hep-th/9607112].
  %%CITATION = doi:10.1103/PhysRevD.55.2373;%%

  \bibitem{Hewson1996} 
  S.~Hewson and M.~Perry,
  ``The Twelve-dimensional super (2+2)-brane,''
  Nucl.\ Phys.\ B {\bf 492}, 249 (1997)
   [hep-th/9612008].
  %%CITATION = doi:10.1016/S0550-3213(97)00120-X, 10.1016/S0550-3213(97)80035-1;%%
     
  \bibitem{Tseytlin1996} 
  A.~A.~Tseytlin,
  ``Type IIB instanton as a wave in twelve-dimensions,''
  Phys.\ Rev.\ Lett.\  {\bf 78}, 1864 (1997)
  [hep-th/9612164].
  %%CITATION = doi:10.1103/PhysRevLett.78.1864;%%
 
   \bibitem{Hurth1997} 
  T.~Hurth, P.~van Nieuwenhuizen, A.~Waldron and C.~Preitschopf,
  ``On a possible new R**2 theory of supergravity,''
  Phys.\ Rev.\ D {\bf 55}, 7593 (1997)
  [hep-th/9702052].
  %%CITATION = doi:10.1103/PhysRevD.55.7593;%%
       
  \bibitem{Khviengia1997} 
  N.~Khviengia, Z.~Khviengia, H.~Lu and C.~N.~Pope,
  ``Towards a field theory of F theory,''
  Class.\ Quant.\ Grav.\  {\bf 15}, 759 (1998)
  [hep-th/9703012].
  %%CITATION = doi:10.1088/0264-9381/15/4/005;%%
  
   \bibitem{Bars1997} 
  I.~Bars and C.~Kounnas,
  ``Theories with two times,''
  Phys.\ Lett.\ B {\bf 402}, 25 (1997)
  [hep-th/9703060].
  %%CITATION = doi:10.1016/S0370-2693(97)00452-8;%%
  
   \bibitem{Nishino1997} 
  H.~Nishino,
  ``N=2 chiral supergravity in (10+2)-dimensions as consistent background for super(2+2)-brane,''
  Phys.\ Lett.\ B {\bf 437}, 303 (1998)
    [hep-th/9706148].
  %%CITATION = doi:10.1016/S0370-2693(98)00924-1;%%

   \bibitem{Rudychev1997} 
  I.~Rudychev, E.~Sezgin and P.~Sundell,
  ``Supersymmetry in dimensions beyond eleven,''
  Nucl.\ Phys.\ Proc.\ Suppl.\  {\bf 68}, 285 (1998)
  [hep-th/9711127].
  %%CITATION = doi:10.1016/S0920-5632(98)00162-5;%%

  \bibitem{Manvelyan1999} 
  R.~Manvelyan and R.~Mkrtchian,
  ``Towards SO(2,10) invariant M theory: MultiLagrangian fields,''
  Mod.\ Phys.\ Lett.\ A {\bf 15}, 747 (2000)
  [hep-th/9907011].
  %%CITATION = doi:10.1142/S0217732300000736, 10.1016/S0217-7323(00)00073-6;%%
  
   \bibitem{Hewson1999} 
  S.~Hewson,
  ``On supergravity in (10,2),''
  hep-th/9908209.
  %%CITATION = HEP-TH/9908209;%%
 
  \bibitem{Vafa1996} 
  C.~Vafa,
  ``Evidence for F theory,''
  Nucl.\ Phys.\ B {\bf 469}, 403 (1996)
  [hep-th/9602022].
  %%CITATION = doi:10.1016/0550-3213(96)00172-1;%%
  
  \bibitem{superspace} 
  S.~J.~Gates, M.~T.~Grisaru, M.~Rocek and W.~Siegel,
  ``Superspace Or One Thousand and One Lessons in Supersymmetry,''
  Front.\ Phys.\  {\bf 58}, 1 (1983)
  [hep-th/0108200];
  %%CITATION = HEP-TH/0108200;%%
  J.~Wess and J.~Bagger,
  ``Supersymmetry and supergravity,''
  Princeton, USA: Univ. Pr. (1992) 
  
   \bibitem{vPF} 
  D.~Z.~Freedman and A.~Van Proeyen,
  ``Supergravity,'' CUP, 2012.
  %%CITATION = INSPIRE-1123253;%%
  
   \bibitem{groupmanifold} 
  L.~Castellani, R.~D'Auria and P.~Fr\'e,
  ``Supergravity and superstrings: A Geometric perspective. 3 Vol.s''
  Singapore, Singapore: World Scientific (1991);
  L.~Castellani, P.~Fr\'e and P.~van Nieuwenhuizen,
  ``A Review of the Group Manifold Approach and Its Application to Conformal Supergravity,''
  Annals Phys.\  {\bf 136}, 398 (1981);
  %%CITATION = doi:10.1016/0003-4916(81)90104-4;%%
  L.~Castellani,
  ``Group geometric methods in supergravity and superstring theories,''
  Int.\ J.\ Mod.\ Phys.\ A {\bf 7}, 1583 (1992).
  %%CITATION = doi:10.1142/S0217751X92000697;%%
  
    \bibitem{Zanelli2005} 
  J.~Zanelli,
  ``Lecture notes on Chern-Simons (super-)gravities. Second edition (February 2008),''
  hep-th/0502193.
  %%CITATION = HEP-TH/0502193;%%
  
  \bibitem{PvN1981} 
  P.~van Nieuwenhuizen,
  ``Supergravity,''
  Phys.\ Rept.\  {\bf 68}, 189 (1981).
  %%CITATION = doi:10.1016/0370-1573(81)90157-5;%%
  
  \bibitem{Castellani2016} 
  L.~Castellani, R.~Catenacci and P.~A.~Grassi,
  ``The Integral Form of Supergravity,''
  JHEP {\bf 1610}, 049 (2016)
  [arXiv:1607.05193 [hep-th]].
  %%CITATION = doi:10.1007/JHEP10(2016)049;%%
  
  \bibitem{MDM} 
  S.~W.~MacDowell and F.~Mansouri,
  ``Unified Geometric Theory of Gravity and Supergravity,''
  Phys.\ Rev.\ Lett.\  {\bf 38}, 739 (1977)
  [Erratum-ibid.\  {\bf 38}, 1376 (1977)].
  %%CITATION = PRLTA,38,739;%%

 \bibitem{Cham} 
  A.~H.~Chamseddine and P.~C.~West,
  ``Supergravity as a Gauge Theory of Supersymmetry,''
  Nucl.\ Phys.\ B {\bf 129}, 39 (1977).
  %%CITATION = doi:10.1016/0550-3213(77)90018-9;%%
  
  \bibitem{PvN2} 
  P.~van Nieuwenhuizen,
  ``Supergravity as a Yang-Mills theory,''
  In *'t Hooft, G. (ed.): 50 years of Yang-Mills theory* 433-456
  [hep-th/0408137].
  %%CITATION = HEP-TH/0408137;%%
   
    \bibitem{Troncoso1998} 
  R.~Troncoso and J.~Zanelli,
  ``Gauge supergravities for all odd dimensions,''
  Int.\ J.\ Theor.\ Phys.\  {\bf 38}, 1181 (1999)
  [hep-th/9807029].
  %%CITATION = doi:10.1023/A:1026614631617;%%
   
  \bibitem{Castellani2013} 
  L.~Castellani,
  ``$OSp(1|4)$ supergravity and its noncommutative extension,''
  Phys.\ Rev.\ D {\bf 88}, no. 2, 025022 (2013)
  [arXiv:1301.1642 [hep-th]].
  %%CITATION = doi:10.1103/PhysRevD.88.025022;%%
  
  
  
  
  
\end{thebibliography}
\end{document}